\documentclass{article}
\usepackage[utf8]{inputenc}
\usepackage{mathtools}
\usepackage{authblk}
\usepackage{fullpage}
\usepackage{graphicx}
\graphicspath{{./figures/}}
\usepackage{subcaption}
\usepackage{amsmath}
\usepackage{amssymb}
\usepackage{textcomp}
\usepackage{multirow}
\usepackage[dvipsnames]{xcolor}
\usepackage{booktabs}
\usepackage{hyperref}
\usepackage[backend=bibtex,casechanger=auto,maxnames=4,minnames=1,doi=true,url=false,eprint=false,citestyle=numeric-comp,sorting=none]{biblatex}
\DeclareFieldFormat[article,inbook,incollection,inproceedings,patent,unpublished]{titlecase:title}{\MakeSentenceCase*{#1}}
\title{Gaussian Process Regression-based Knowledge Distillation Framework for Simultaneous Prediction of Physical and Mechanical Properties of Epoxy Polymers}
\author[a]{Sindu B.S.}
\author[b]{Jan Hamaekers}
\affil[a]{{\small Special and Multifunctional Structures Laboratory, CSIR-Structural Engineering Research Centre, Taramani, Chennai, Tamil Nadu, India - 600113.}}
\affil[b]{{\small Fraunhofer Institute for Algorithms and Scientific Computing SCAI, Schloss Birlinghoven, \mbox{53757 Sankt Augustin}, Germany}}
\date{\today}
\begin{document}
\maketitle
\begin{abstract}
Epoxy polymers are widely used due to their multifunctional properties, but machine learning (ML) applications remain limited owing to their complex 3D molecular structure, multi-component nature, and lack of curated datasets. Existing ML studies are largely restricted to simulation data, specific properties, or narrow constituent ranges. To address these limitations, we developed an informed Gaussian Process Regression-based Knowledge Distillation (GPR-KD) framework for predicting multiple physical (glass transition temperature, density) and mechanical properties (elastic modulus, tensile strength, compressive strength, flexural strength, fracture energy, adhesive strength) of thermoset epoxy polymers. The model was trained on experimental literature data covering diverse monomer classes (9 resins, 40 hardeners). Individual GPR models serve as teacher models capturing nonlinear feature-property relationships, while a unified neural network student model learns distilled knowledge across all properties simultaneously. By encoding the target property as an input feature, the student model leverages cross-property correlations. Molecular-level descriptors extracted from SMILES representations using RDKit create a physics-informed model. The framework combines GPR interpretability and robustness with deep learning scalability and generalization. Comparative analysis demonstrates superior prediction accuracy over conventional ML models. Simultaneous multi-property prediction further improves accuracy through information sharing across correlated properties. The proposed framework enables accelerated design of novel epoxy polymers with tailored properties.

         
\end{abstract}

\section{Introduction}
Epoxy polymers are thermoset materials exhibiting with multi-functional properties such as high strength, excellent adhesion, effective electrical insulation, low shrinkage, chemical and solvent resistance, etc. They are widely used in several industries like aerospace, marine, automotive, infrastructure, electrical and electronics for specific functional requirement. For instance, adhesive strength, fracture toughness and weight are the major requirements for use in aerospace applications \cite{jin2015synthesis}; resistance to moisture, salinity and varied temperature cycles is the major requirement in marine applications; viscosity, low shrinkage and corrosion resistance are the requirements for its application in paint and coatings \cite{jin2015synthesis}; insulation resistance is the major requirement for electrical applications \cite{shundo2022network}; adhesive strength, durability, tensile and fracture properties are the requirements for infrastructural applications \cite{rahman2022application}. Hence, care should be taken to design epoxy polymers with required functionalities so that it can be used for the intended purposes. The basic constituents of epoxy polymers are resin and hardener. When these two compounds undergo an irreversible reaction, a dense, three dimensional network is formed. However, the functionalities and the properties of epoxy polymers depend upon several factors like the constituent resin and hardener types, their composition, curing conditions and degree of polymerization. The conventional trial-and-error-based experimental investigations limit the development of high performance epoxy polymers with multi-functional characteristics.     

A few novel approaches were developed in the recent times to establish constituent-structure-property relationship of epoxy polymers and to predict its properties which can enrich the design process. Results from experimental nanoindentation \cite{zhang2020characterization, passilly2019characterization} tests and scratch induced deformation tests \cite{molero2019scratch} were used to establish a correlation with the intrinsic mechanical properties. Computational molecular dynamics (MD) simulations were performed to determine the mechanical, physical and transport properties of epoxy polymers \cite{sindu2022atomistic}. MD simulations using Reactive Interface Force Field (IFF-R) were used to determine the properties of epoxy polymers when subjected to large deformations \cite{odegard2021molecular}; to get the experimental results that can be obtained using Differential Scanning Calorimetry (DSC)~\cite{okabe2016molecular} by utilizing information such as heat of formation and activation energy. MD simulations were also used to identify the influence of certain molecular interactions on mechanical and transport properties of polymers \cite{okabe2016molecular, sindu2015evaluation}, to correlate the functionality of monomers (di-, tri-, tetra-) with the mechanical properties like yield stress, elastic modulus and Poisson's ratio \cite{radue2018comparing}. 

ML-based approaches are being widely used for design, optimization and property prediction of homopolymers and copolymers due to the availability of large databanks \cite{amamoto2022data, zeng2018graph, kuenneth2021polymer, tao2021benchmarking}. However, their use in thermoset epoxy polymers is very limited due to the involvement of two or more constituent ingredients, complex three-dimensional structure and lack of large curated datasets. Recently, attempts are being made for composition optimization, property prediction and design of high performance epoxy polymers using ML-based approaches. Results from MD simulations in conjunction with Neural Networks (NNs) was used to optimize the composition of epoxy polymers with performance characteristics like elastic modulus, tensile strength, elongation at break, glass transition temperature ($\mathrm{Tg}$) \cite{jin2020composition} and self-healing properties \cite{luo2021properties}. The composition of shape memory epoxy polymers (SMP) was also optimized using NNs using the results from experimental investigations \cite{liu2022performance}. Unified ML model based on Support Vector Regression (SVR) was proposed to predict the $\mathrm{Tg}$ of homo-, hetero- and cross linked epoxy- polymers based on the constituent monomers and its descriptors \cite{higuchi2019prediction}. ML ensemble model based on Gradient Boosting Regression (GBR) and Kernel Ridge Regression (KRR) was used to predict the $\mathrm{Tg}$ of epoxy polymers by correlating the molecular descriptors of resin (16 types) and hardener (19 types) with experimental (94 combinations) DMTA measurements and the most important descriptors affecting the $\mathrm{Tg}$ were further identified using Lasso regression \cite{meier2022modeling}. Materials genome approach in conjunction with attention- and gate-augmented graph convolutional networks, multilayer perceptrons and transfer learning was used to identify the gene structures responsible for different properties and to design epoxy polymers with high strength, modulus and toughness \cite{hu2022machine}. Similarly, ML-based approach was developed to predict the elastic modulus and yield strength of epoxy polymers using the basic structural features of monomers, thereby, enabling feature-based prediction of their properties \cite{sindu2025feature}.  ML-based convolutional model was employed for discovery of new thermoset SMPs with high recovery stress from a newly constituted compositional space \cite{yan2021machine, yan2021drug}.

Though several attempts have been made to use ML for design of epoxy polymer or prediction of properties, most of the approaches seem to be limited as the data for the model are obtained from computational simulations or the model has been developed to predict a specific property of epoxy polymers or with very limited set of constituents. With this in mind, we develop a ML model which can be used to predict multiple physical (glass transition temperature and density) and mechanical (elastic modulus, tensile strength, compressive strength, flexural strength, fracture energy and adhesive strength) properties of two-component, thermoset epoxy polymers merely with the help of its constituent ingredients (type of resin and hardener and its proportion). We develop a GPR-based knowledge distillation framework (GPR-KD) for this purpose in which individual GPR models are trained as teacher models for each target property, and the the knowledge learned from these teachers is distilled into a unified student model for property prediction. The constituent resin and hardener are first encoded using a label encoding scheme to numerically represent their chemical identities. These encoded variables are then used together with the relevant process parameters as inputs to the model, allowing it to learn correlations between material composition, processing conditions, and the resulting physical and mechanical properties (cf.\ Section~\ref{sec:2.1}). We then convert it into an informed ML model \cite{von2021informed} by including information about the intrinsic features of the monomeric constituents extracted from an open source, cheminformatics tool (cf.\ Section~\ref{sec:2.2}). The data for training the model was collected from several experimental data from literature. We also compared the performance of the model with other conventional approaches and found that our model is able to predict the properties more accurately (cf.\ Section~\ref{sec:3.1}). The major advantage of this model is that a single model can be used to predict multiple physical and mechanical properties of epoxy polymers. The model also extracts the benefits of learning from one another as it has been trained with different properties. This has also been witnessed by improved prediction accuracy while multiple properties are predicted together (cf.\ Section~\ref{sec:3.2}). The model thus paves the path for designing novel epoxy polymers with exceptional properties which would only be possible through laborious experimental trials.  

\section{ML model for prediction of properties} \label{sec:2}
\begin{figure}[hbt!]
    \centering
    \begin{subfigure}{0.6\textwidth}
        \includegraphics [width=0.95\textwidth]{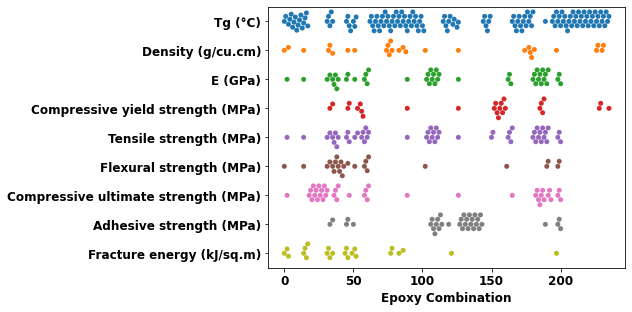}
        \caption{}\label{fig:1a}
    \end{subfigure}
    \begin{subfigure}{0.4\textwidth}
        \includegraphics [width=0.95\textwidth]{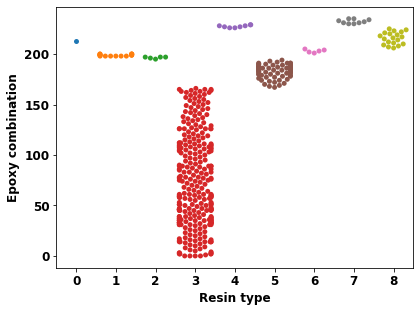}
        \caption{}\label{fig:1b}
    \end{subfigure}
    \begin{subfigure}{0.4\textwidth}
        \includegraphics [width=0.95\textwidth]{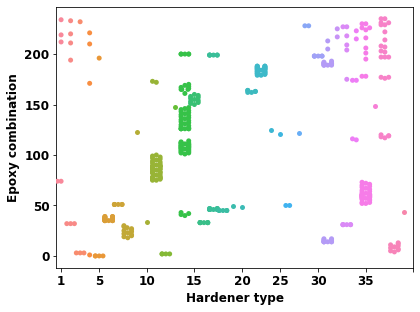}
        \caption{}\label{fig:1c}
    \end{subfigure}
    \begin{subfigure}{0.95\textwidth}
        \includegraphics [width=0.95\textwidth]{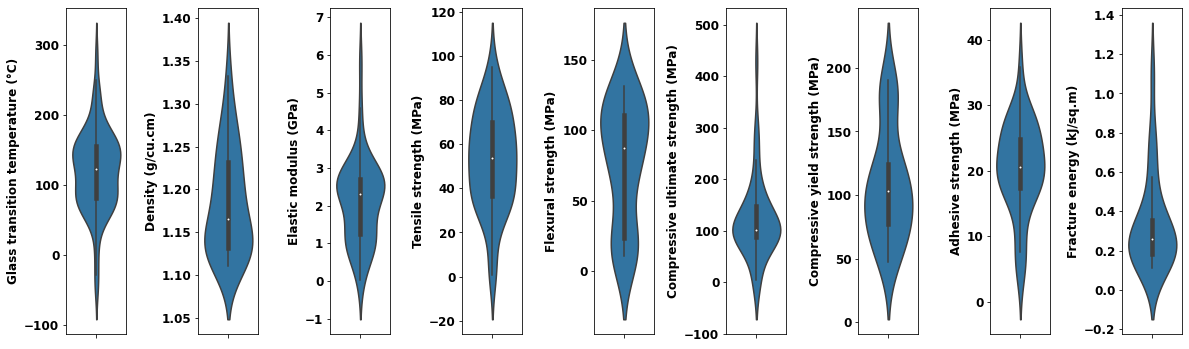}
        \caption{}\label{fig:1d}
    \end{subfigure}
    \caption{Data collected from literature (used for training the ML model): (\subref{fig:1a}) Distribution of datapoints for each property across different epoxy combinations (236 combinations in total); Distribution of data across individual (\subref{fig:1b}) resin classes and (\subref{fig:1c}) hardener classes; (\subref{fig:1d}) Range of individual physical and mechanical properties.}
    \label{fig:1}
\end{figure}

In order to predict the physical and mechanical properties of epoxy polymers, we develop a GPR-based knowledge distillation framework. The data for training the ML model is provided from the experimental results in the literature. A total of 236 data points spread across several properties such as glass transition temperature ($\mathrm{Tg}$), density, elastic modulus, compressive strength, tensile strength, flexural strength and adhesive strength for different types of epoxy polymers are collected for this purpose \cite{garcia2007mechanical, gonzalez2010influence, knorr2012glass, bellenger1988packing, pineda2017comparative, prozonic2012effect, lee2000role, pruksawan2019prediction, prolongo2006comparative, jang2012particle, selby1975fracture, kinloch2005mechanics, aziz2011study, liu2007solid, vanlandingham1999relationships, elmahdy2021effect, behzadi2005yielding, sun2008mechanical, littell2008measurement,
jordan2008mechanical, jakubinek2018nanoreinforced, fard2012characterization, chen2002tension, rostamiyan2015experimental, kochergin2010properties, morrill2004prediction}. The data spans across diverse resin (9) and hardener (40) classes; several process parameters (stoichiometric ratio between the resin and hardener, curing temperature) and test parameters (strain rate, test temperature). The distribution of datapoints across different epoxy combinations and the range of individual physical and mechanical properties collected from literature is presented in Figure \ref{fig:1}.  

\subsection{Basic architecture of the model} \label{sec:2.1}
The basic input used for prediction of epoxy polymers properties using the proposed GPR-KD framework are the constituent ingredient (resin and hardener) types, their proportion and other process parameters. Firstly, we convert the categorical data (resin type, hardener type and property to be predicted) into numerical labels using label encoder. Then, this encoded information along with the molecular weight of individual monomers, process parameters and test parameters are selected as input features for the model.
\begin{figure}[hbt!]
    \centering
    \includegraphics[width=0.98\textwidth]{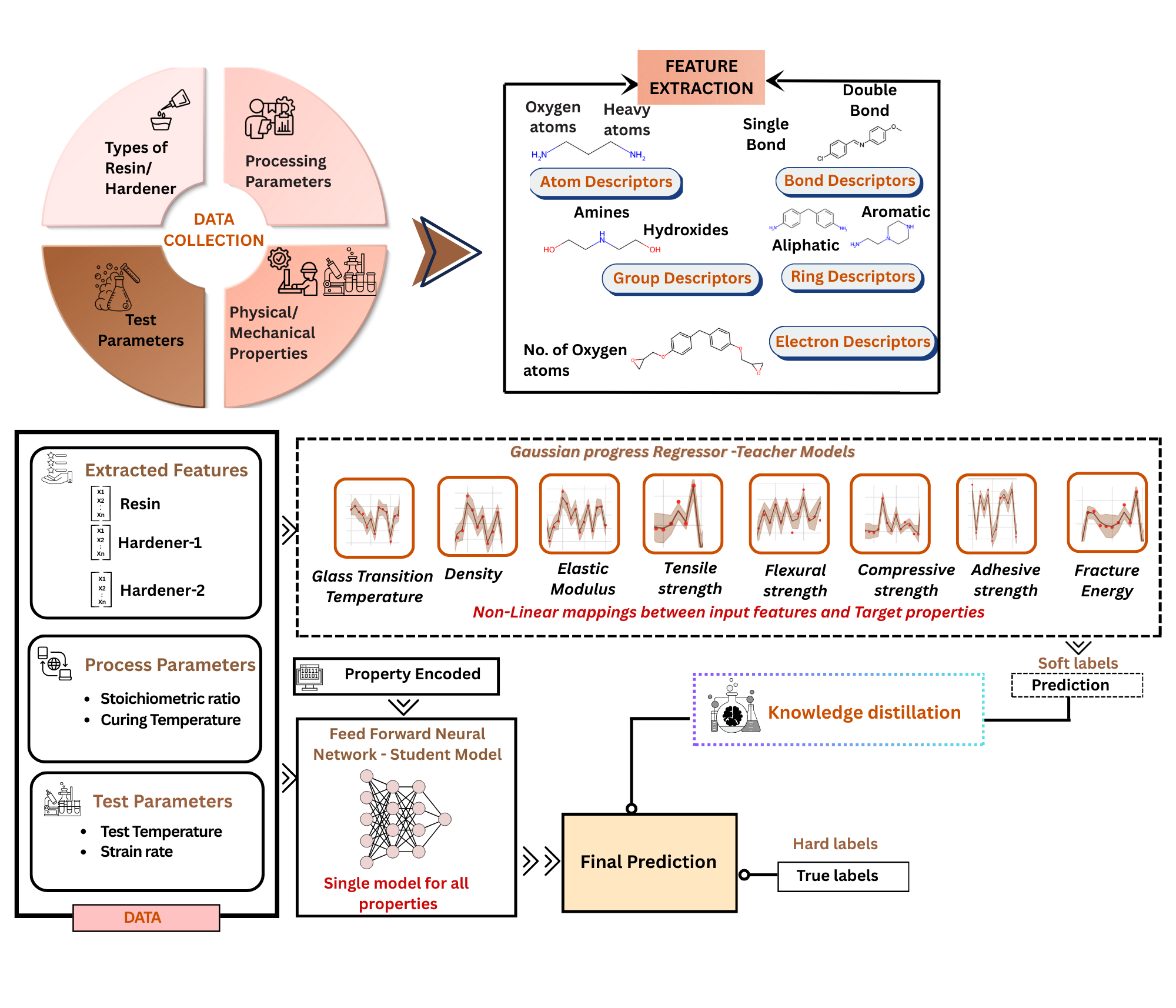}
    \caption{Architecture of the GPR-KD framework used in this study.}
    \label{fig:2}
\end{figure}

For each target property, an independent GPR model is developed and employed as a high-fidelity teacher within the proposed knowledge distillation framework. For training each teacher model, only the subset of the dataset comprising data points associated with the target property is used. The dataset is then normalized to a uniform scale so as to remove bias arising from differences in feature scales. The available data are partitioned into training and testing sets using an 80:20 split. Then, hyperparameter optimization is carried out on training dataset using  five-fold cross-validation. A grid search strategy is employed to explore an extensive kernel space constructed from linear (DotProduct), radial basis function (RBF), Matérn, and constant kernels, including their additive combinations, together with the noise regularization parameter. Model selection is based on minimization of the mean absolute error (MAE), ensuring robust performance across folds while avoiding overfitting. Once the optimal hyperparameter configuration is identified, the best-performing GPR model is retained for final evaluation and knowledge transfer. The trained teacher model is used to generate predictions on the full normalized input space to produce property-specific soft targets for subsequent knowledge distillation. These GPR teacher models thus provide accurate non-linear mappings between the input features and target properties, yielding physically consistent predictive responses that are well suited for guiding the training of a unified student neural network.

The student model is implemented as a fully connected feed-forward neural network with an input layer, two hidden layers, and an output layer. The input to the student network is formed by concatenating the normalized feature vector with a one-hot encoded representation of the property to be predicted. This explicit conditioning enables the student network to distinguish among different target properties while sharing a common set of network parameters, thereby allowing simultaneous learning of multiple properties within a single model architecture. The concatenated input is passed sequentially through two hidden layers, where weights and biases are learned through non-linear activation using the rectified linear unit (ReLU) function. Through these hidden layers, the model progressively transforms the raw input into a higher-dimensional latent representation that captures coupled constituent–process–property relationships, which are not directly apparent from the original feature space. The final output layer maps this latent representation to a single scalar corresponding to the predicted epoxy polymers property. The model is trained using PyTorch Lightning \cite{1370013168774120069}. During training, a knowledge distillation loss function is employed, defined as a weighted sum of the mean squared error between the student predictions and the GPR teacher predictions (soft targets) and the mean squared error with respect to the true experimental values.
A weighting factor of $\alpha=0.7$ is used, assigning higher importance to matching the teacher predictions while retaining direct supervision from experimental data. Accordingly, the overall loss function reads as
\begin{equation*}
\mathcal{L}_{\text{KD}} =
\alpha \, \mathrm{MSE}\!\left(\hat{y}, y_{\text{teacher}}\right)
+
(1-\alpha) \, \mathrm{MSE}\!\left(\hat{y}, y_{\text{true}}\right).
\end{equation*}
A schematic overview of the newly developed GPR-KD framework is given in Figure~\ref{fig:2}


This distillation strategy enables the student model to inherit the smooth, physically consistent response characteristics of the GPR teachers while maintaining direct consistency with measured data, resulting in a compact and efficient surrogate model suitable for large-scale parametric studies and rapid property prediction. Training of the student neural network is carried out using the Adam optimizer with a fixed learning rate of $10^{-3}$, and a mini-batch size of 32, for a total of 5000 training epochs.

\subsection{Informed GPR-KD framework} \label{sec:2.2}

\begin{figure}[hbt!]
    \centering
    \begin{subfigure}{0.48\textwidth}
        \includegraphics [width=0.98\textwidth]{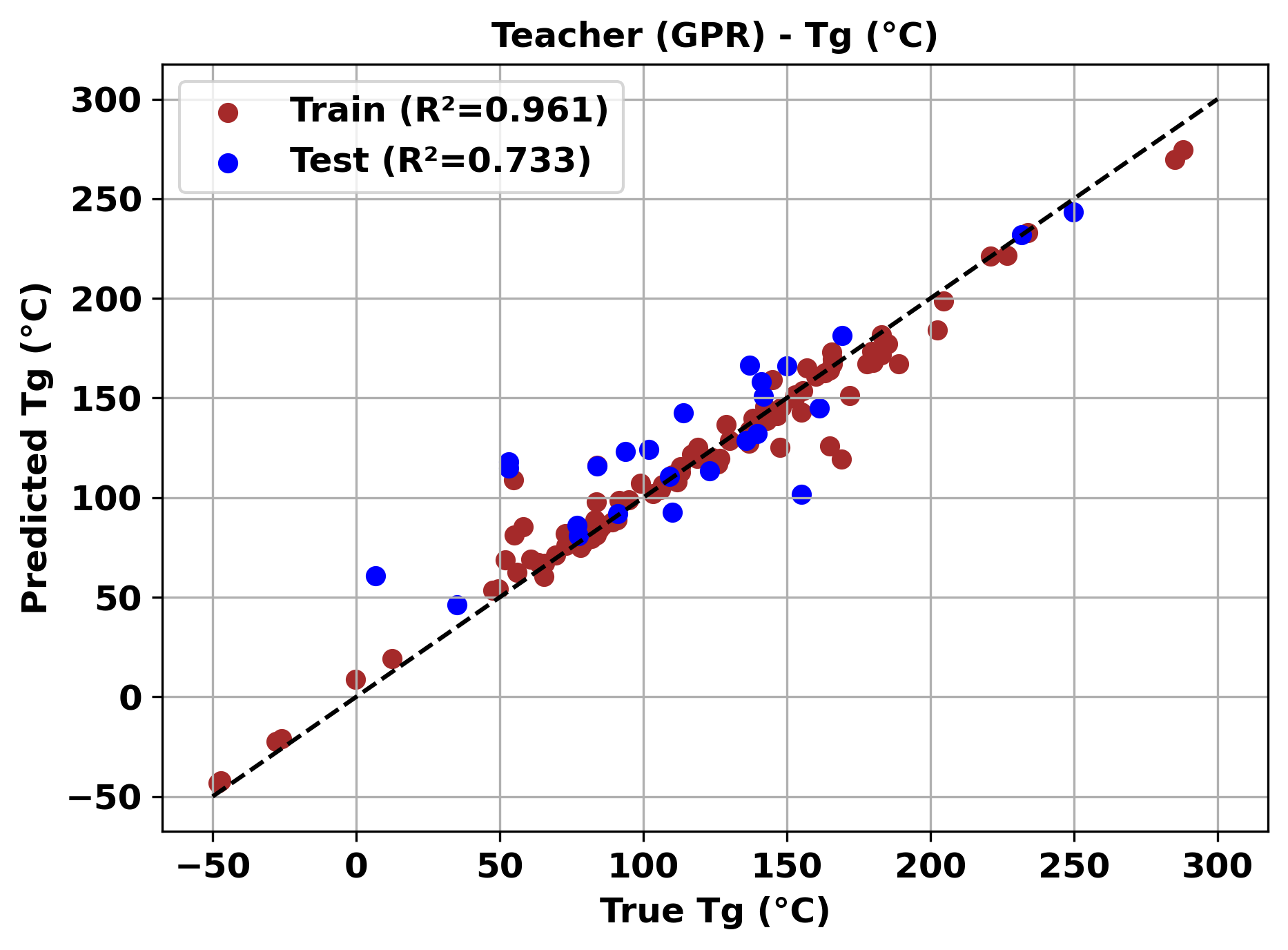}
    \end{subfigure}
    \begin{subfigure}{0.48\textwidth}
        \includegraphics [width=0.98\textwidth]{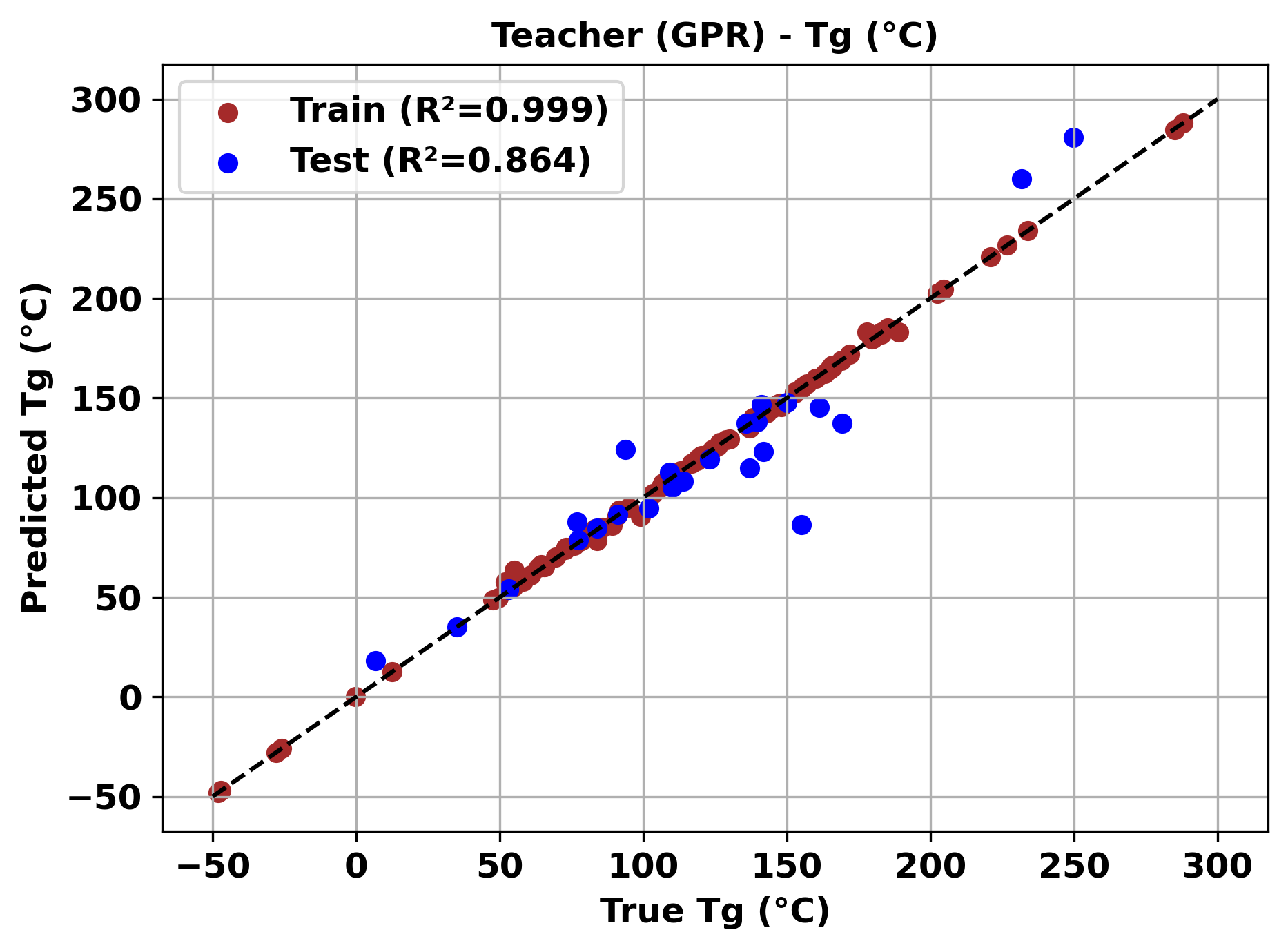}
    \end{subfigure}
    \begin{subfigure}{0.48\textwidth}
        \includegraphics [width=0.98\textwidth]{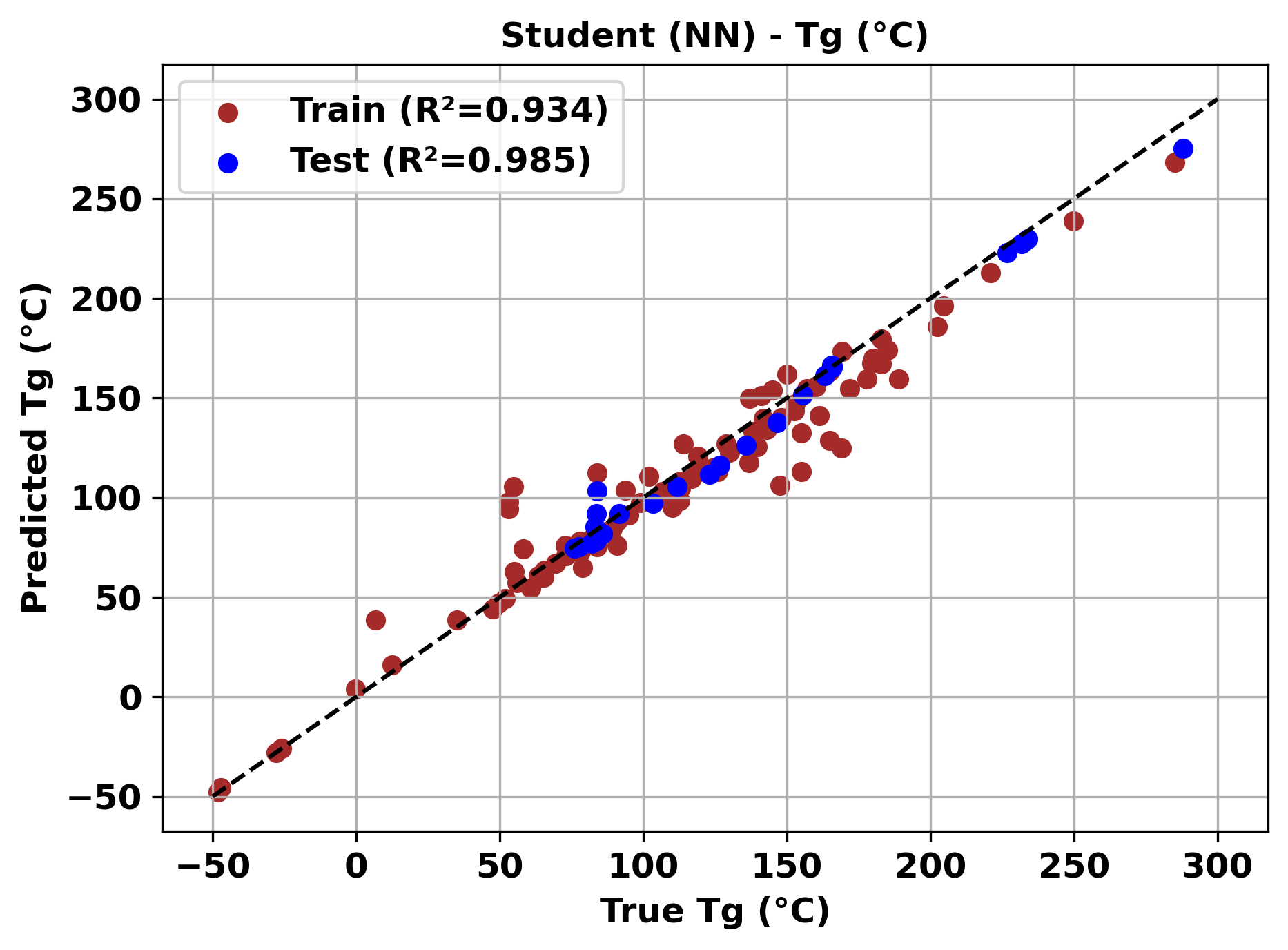}
        \caption{}\label{fig:3a}
    \end{subfigure}
    \begin{subfigure}{0.48\textwidth}
        \includegraphics [width=0.98\textwidth]{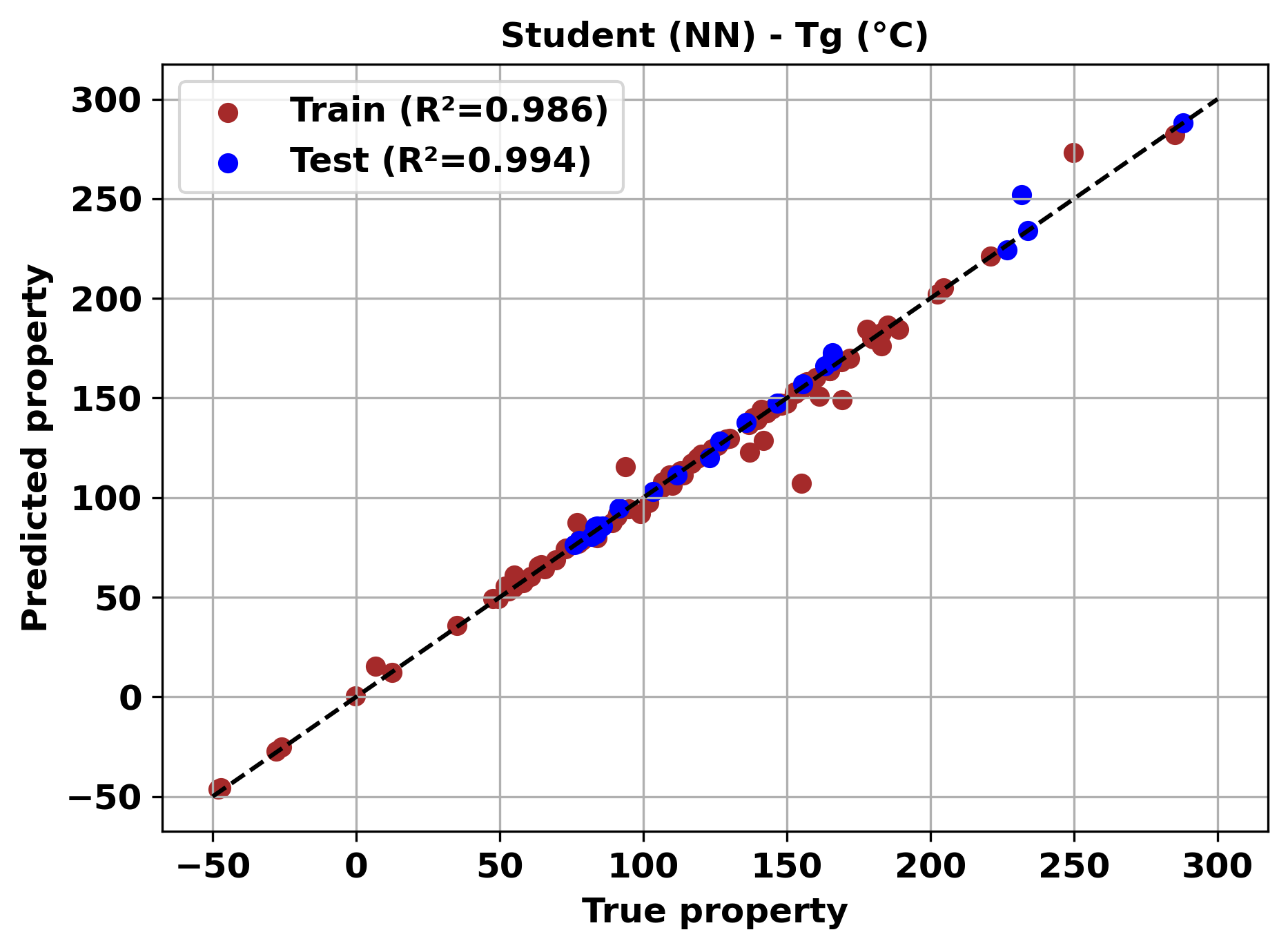}
        \caption{}\label{fig:3b}
    \end{subfigure}
\caption{Prediction accuracy of the proposed (\subref{fig:3a}) Knowledge Distillation Framework; (\subref{fig:3b}) Informed Knowledge Distillation Framework.}
\label{fig:3}
\end{figure}
In order to further enhance the predictive capability of the proposed framework as a physics-informed model, intrinsic molecular-level features of the constituent monomers are incorporated. Each resin and hardener monomer is represented by its unique SMILES string \cite{weininger1988smiles}, which provides an ASCII-based representation of the molecular structure. From these SMILES strings, atomistic and structural descriptors of the resin and hardener molecules are extracted using the open-source cheminformatics tool RDKit \cite{RDKit}. A total of 28 features which include the following descriptors are extracted from RDKit using a python script which include: 
\begin{itemize}
    \item Molecular weight
    \item Atom Descriptors (type and count of atoms including heavy atoms)
    \item Bond Descriptors (single/double/triple)
    \item Group count (NH/OH, SP3 fractions)
    \item Ring counts (aromatic/saturated/aliphatic and heterocycles/carbocycles)
    \item Electron descriptors (No.\ of hydrogen acceptors/donors, radical/valence electrons)
\end{itemize}
In cases where two types of hardeners are present in the formulation, descriptors corresponding to both hardeners are included. In the informed GPR-KD framework, these molecular descriptors replace the abstract categorical representations used in the uninformed architecture, thereby embedding the resin and hardener molecules directly in their physically meaningful feature space. Since the extracted features span several orders of magnitude, all molecular descriptors are normalized between 0 and 1 using a Min–Max scaler to eliminate bias arising from scale differences. The improvement in prediction accuracy, particularly for the glass transition temperature, achieved by embedding the constituent molecules in their feature space is demonstrated in Figure~\ref{fig:3}.

\section{Performance of the informed GPR-KD framework} \label{sec:3}
The performance of the developed informed GPR–KD framework in predicting various physical properties (glass transition temperature and density) and mechanical properties (elastic modulus, tensile strength, flexural strength, compressive yield strength, compressive ultimate strength, adhesive strength, and fracture energy) of epoxy polymers is evaluated using the dataset shown in Figure~\ref{fig:1}. The results of this evaluation are presented and discussed in this section.
\subsection{Comparison with conventional methods} \label{sec:3.1}
\begin{table}[hbt!]
    \caption{Hyperparameter search space for different regression models.}    
    \centering
    \begin{tabular}{lll}
            \hline
            Model & Hyperparameter & Search space\\  
            \hline
            PLS Regression & Number of components & 1 to 13\\
                           & Tolerance & $10^{-4}$, $10^{-6}$, $10^{-8}$, $10^{-10}$\\ 
            Ridge Regression & $\alpha$ & logspace(-10 to 2)\\
                             & Solver & svd, cholesky, lsqr, sparse\_cg,\\
                             & & sag, saga, lbfgs\\
            Kernel Ridge Regression & $\alpha$ & logspace(-10 to 6)\\
                                    & Kernel & rbf, linear, poly, sigmoid\\
            Random Forest & Maximum depth & [1, 2, 3]\\
            Gradient Boosting Regression & Learning rate & logspace(-6 to -1)\\
                              & Maximum depth & [1, 2, 3]\\
            k-Nearest Neighbour & Number of neighbours & 1 to 7\\
                              & Weights & Uniform, Distance\\
            Gaussian Process Regression & $\alpha$ & logspace(-12 to -1)\\
            \hline
            \end{tabular}
    \label{tab:1}
\end{table}
\begin{figure}[hbt!]
    \centering
    \includegraphics[width=0.95\textwidth]{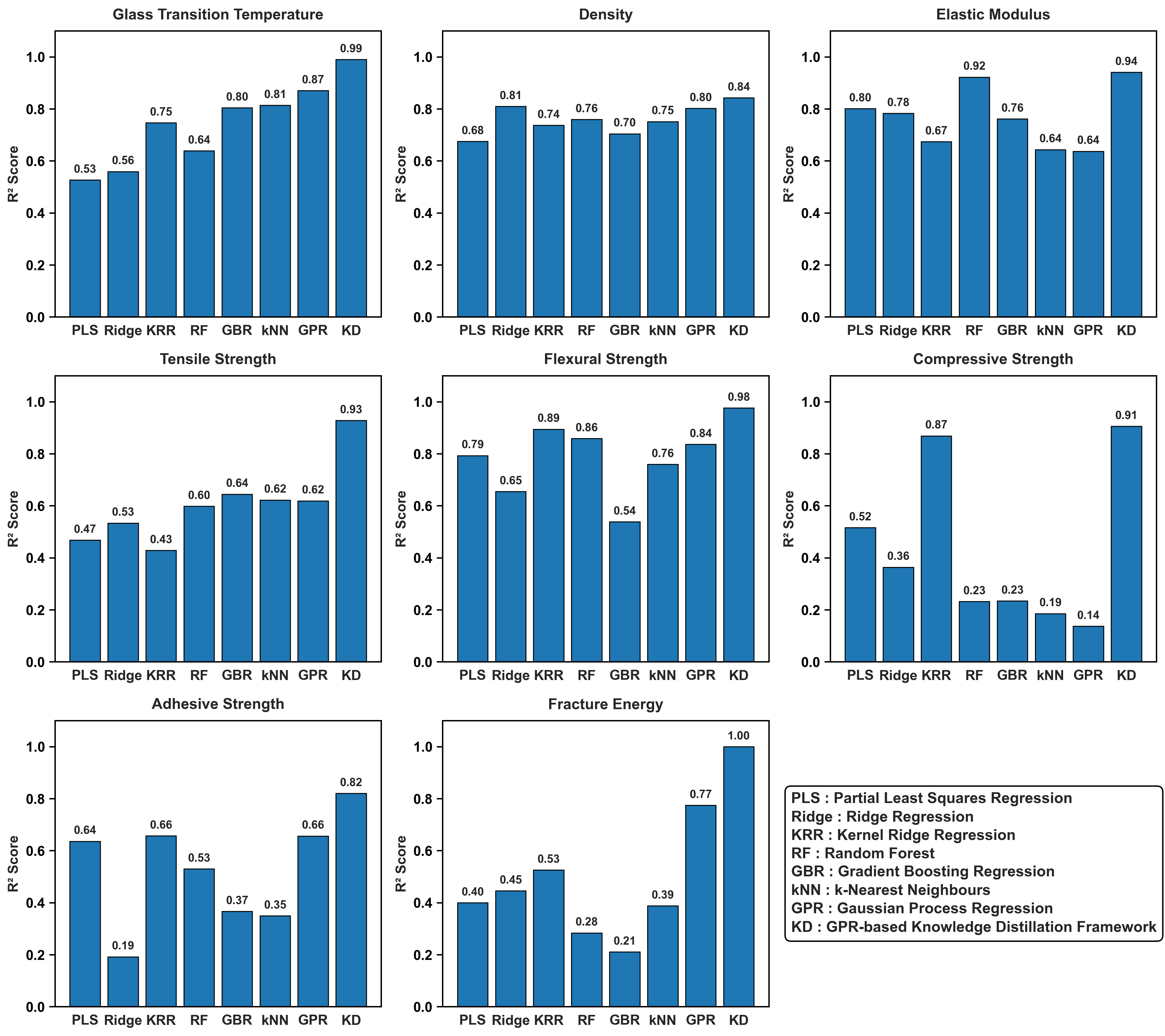}
    \caption{Prediction accuracy (in terms of $R^2$ score) of different physical and mechanical properties using proposed GPR-based knowledge distillation framework and conventional models.}
    \label{fig:4}
\end{figure}
 The predictive capability of the informed GPR–KD framework is assessed by comparing its performance with several conventional ML models including Partial Least Squares (PLS) Regression, Ridge Regression (RR), Kernel Ridge Regression (KRR), Random Forest (RF), Gradient Boosting Regression (GBR), k-Nearest Neighbours (kNN) and  Gaussian Process Regression (GPR). For all models, the input features comprise of molecular-level descriptors (as described in Section~\ref{sec:2.2}) of the constituent resin and hardener(s) extracted from cheminformatics tool along with relevant process and testing parameters, while the target epoxy property serves as the output variable. Prior to model training, all continuous input features are normalized using Min–Max scaling to eliminate bias arising from differences in feature magnitudes and to ensure stable optimization across learning algorithms. For each target property, the available data are partitioned into training and testing sets using an 80:20 split. Model development and hyperparameter optimization are performed exclusively on the training set.
 Each conventional model involves a set of hyperparameters that significantly influence its predictive performance. Accordingly, hyperparameter optimization is carried out using a grid search strategy combined with five-fold cross-validation on the training data. The hyperparameter search spaces for the different models are defined individually for each algorithm (presented in Table~\ref{tab:1}). Model selection is based on minimization of the mean absolute error, while the generalization capability of the optimized models is evaluated using the held-out test set. 
Figure \ref{fig:4} shows the prediction accuracy of informed GPR-KD framework in comparison to conventional ML models. It can be found that the informed GPR-KD model consistently exhibits higher $R^2$ scores between the true and predicted property values than the conventional ML models for all the properties considered.

\subsection{Informed GPR-KD framework for simultaneous prediction of physical and mechanical properties} \label{sec:3.2}
\begin{figure}[hbt!]
    \centering
    \includegraphics[width=0.9\textwidth]{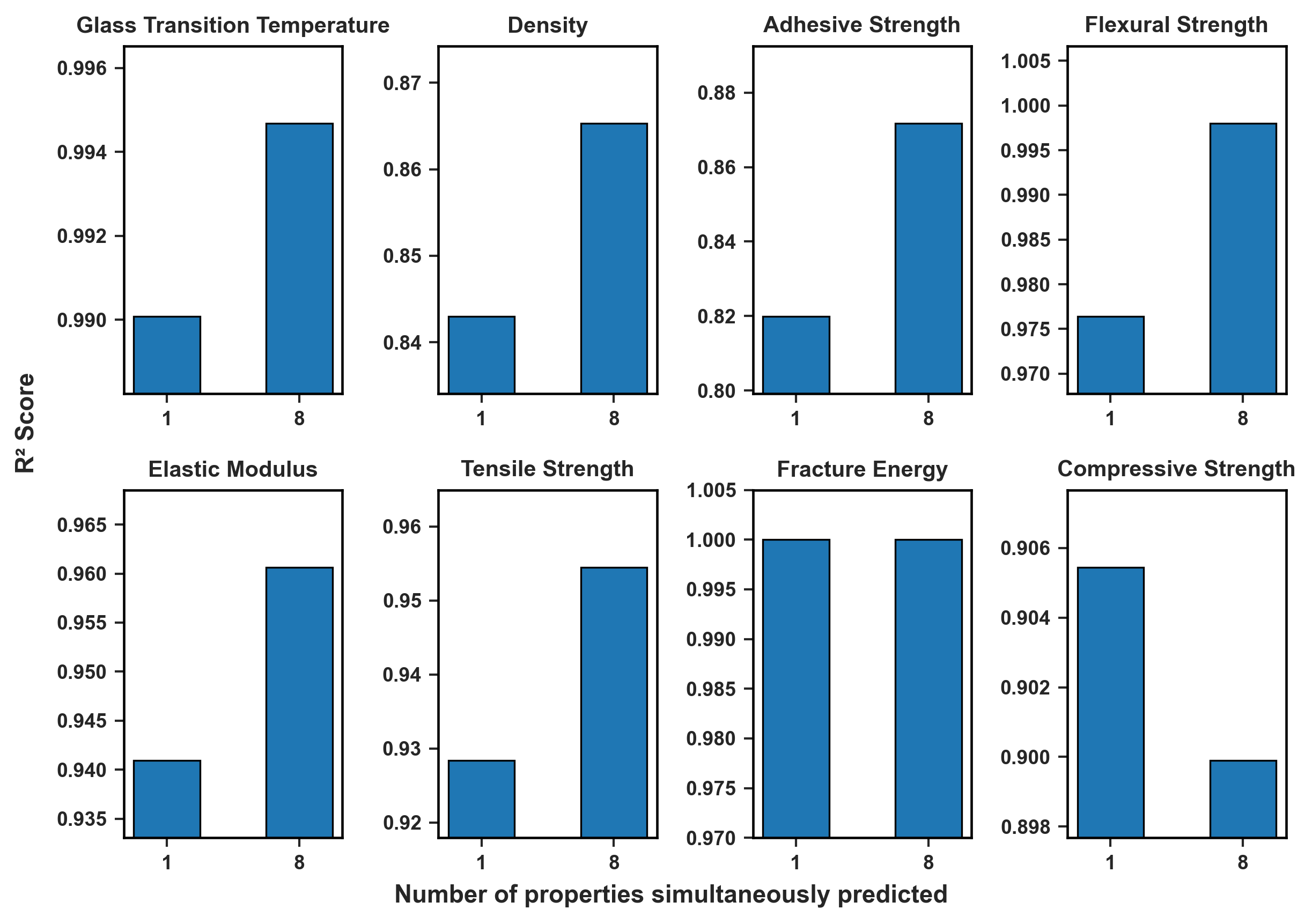}
    \caption{Prediction accuracy (in terms of $R^2$ Score) during simultaneous prediction of multiple properties.}
    \label{fig:5}
\end{figure}

\begin{figure}[hbt!]
    \centering
    \begin{subfigure}{0.325\textwidth}
        \includegraphics [width=0.99\textwidth]{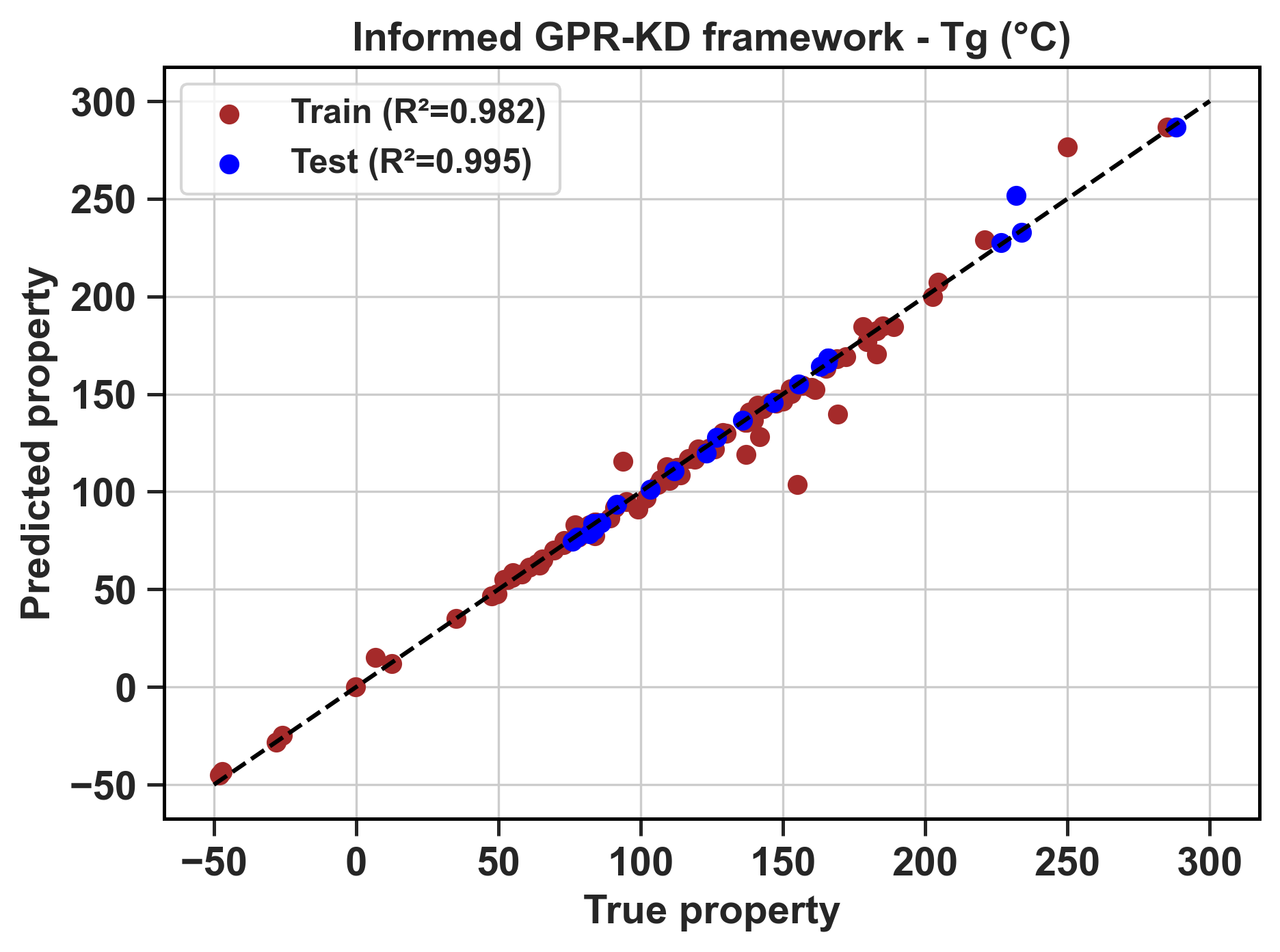}
        \caption{}\label{fig:6a}
    \end{subfigure}
        \begin{subfigure}{0.325\textwidth}
        \includegraphics [width=0.99\textwidth]{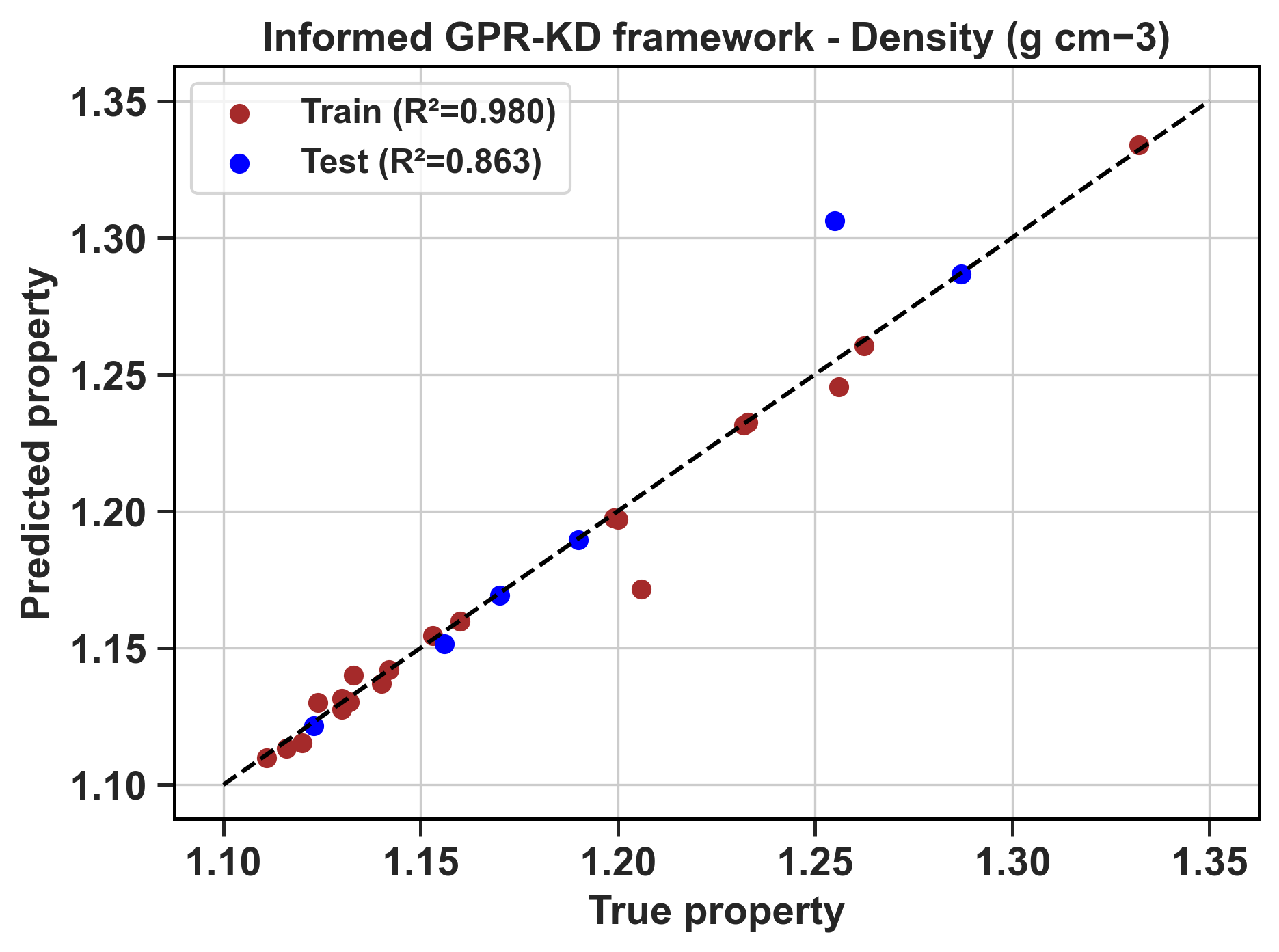}
        \caption{}\label{fig:6b}
    \end{subfigure}
        \begin{subfigure}{0.325\textwidth}
       \includegraphics [width=0.99\textwidth]{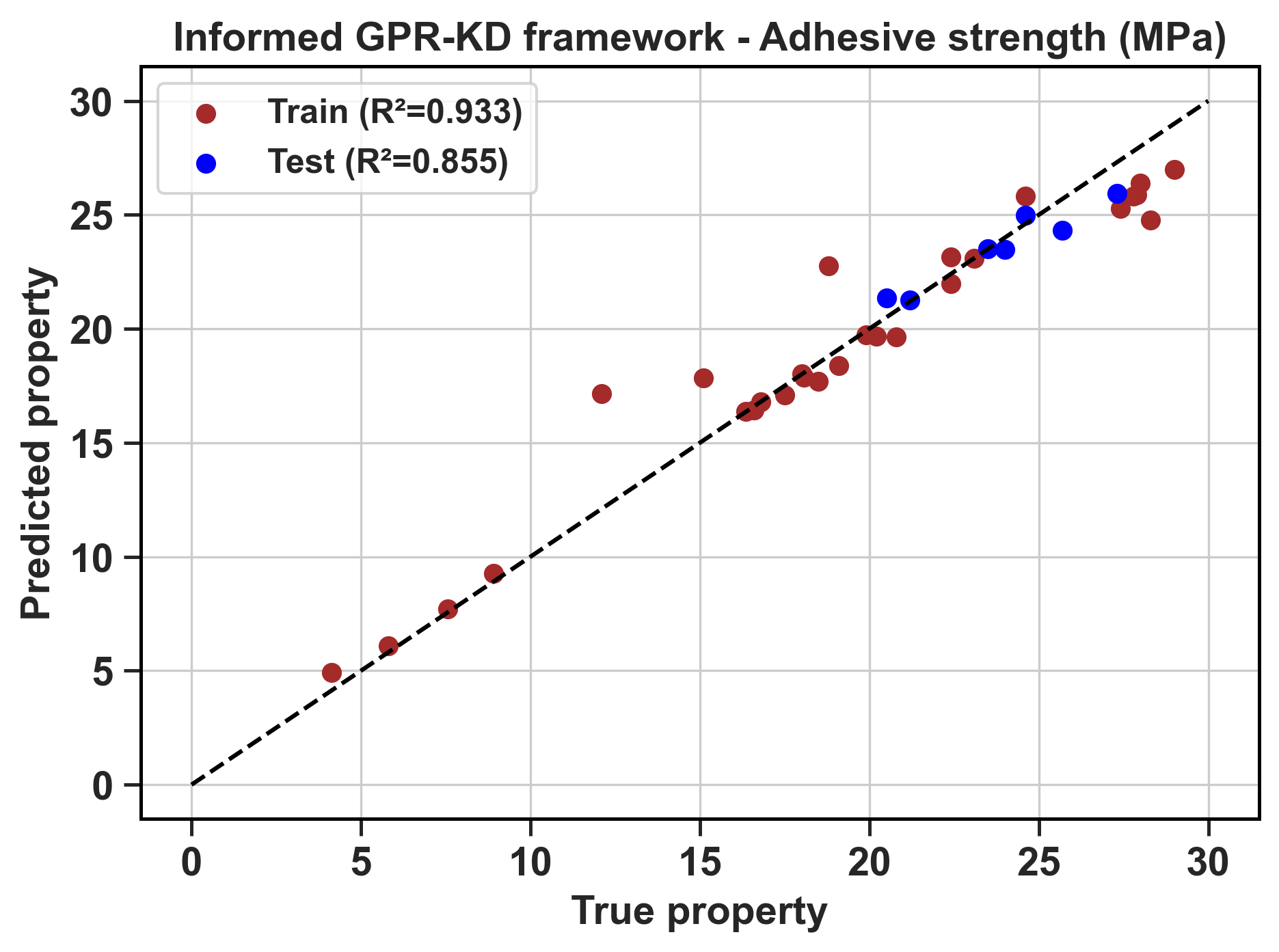}
       \caption{}\label{fig:6c}
    \end{subfigure}
        \begin{subfigure}{0.325\textwidth}
        \includegraphics [width=0.99\textwidth]{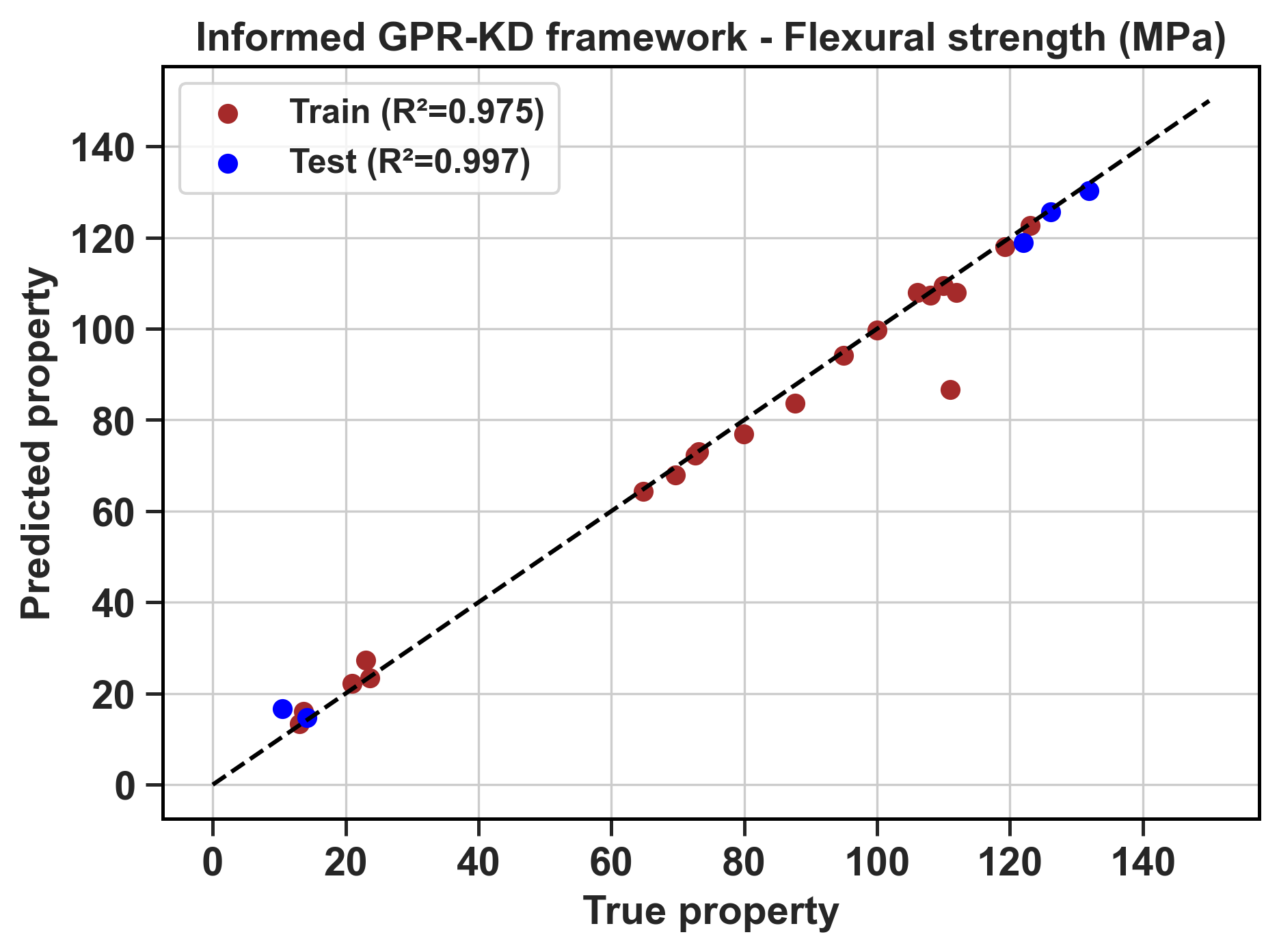}
        \caption{}\label{fig:6d}
    \end{subfigure}
    \begin{subfigure}{0.325\textwidth}
        \includegraphics [width=0.99\textwidth]{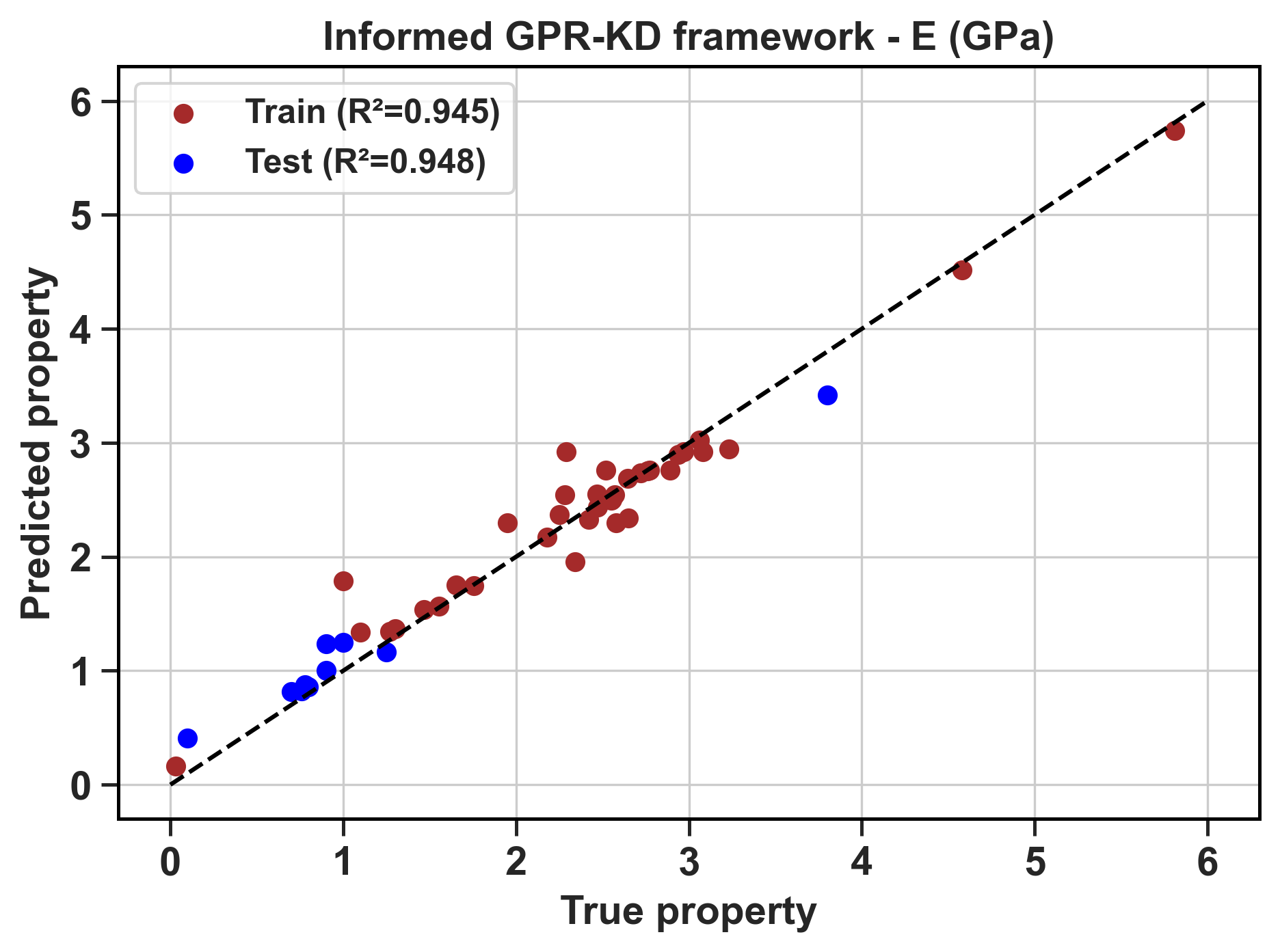}
        \caption{}\label{fig:6e}
    \end{subfigure}
    \begin{subfigure}{0.325\textwidth}
        \includegraphics [width=0.99\textwidth]{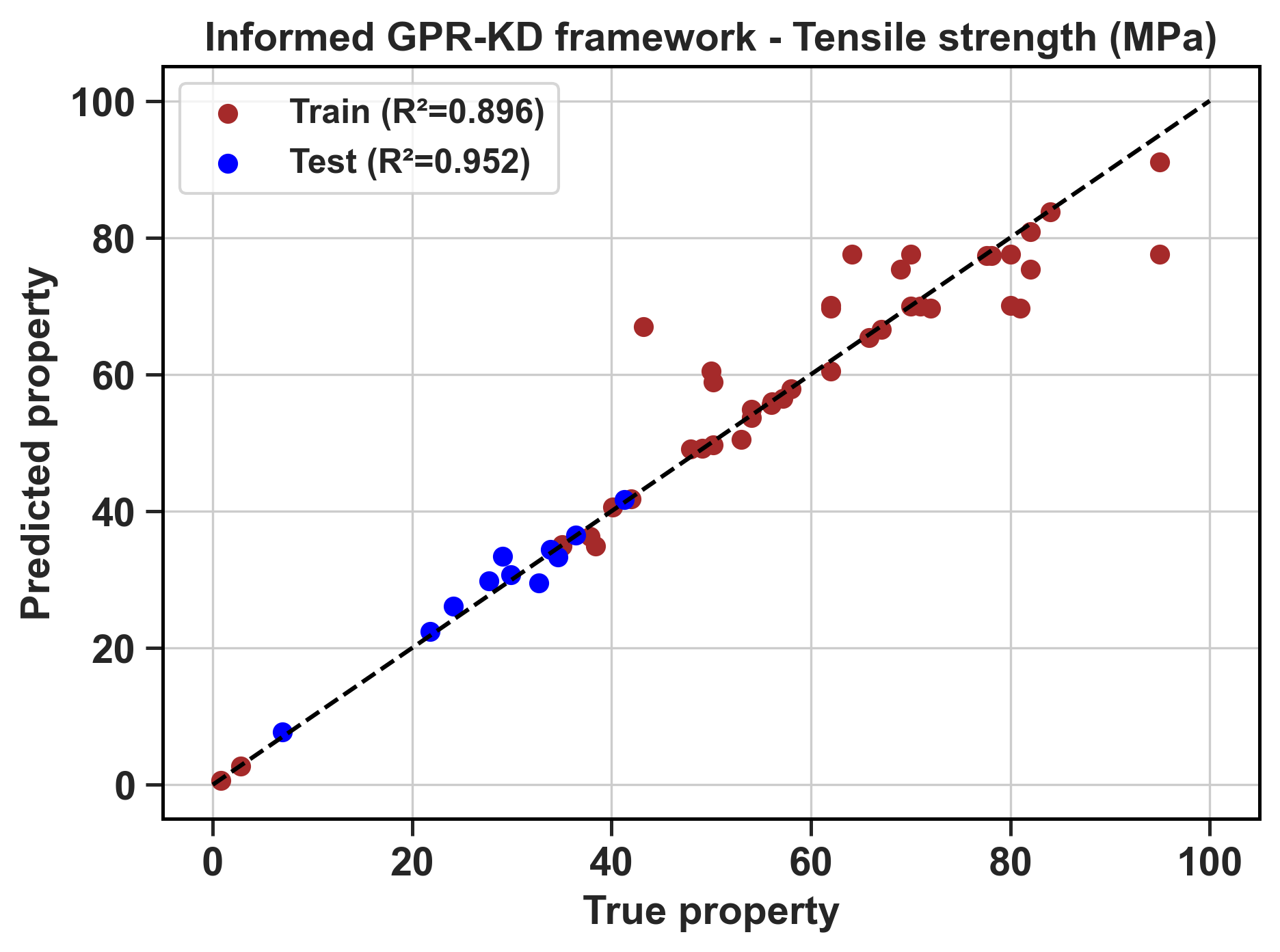}
        \caption{}\label{fig:6f}
    \end{subfigure}
    \begin{subfigure}{0.325\textwidth}
        \includegraphics [width=0.99\textwidth]{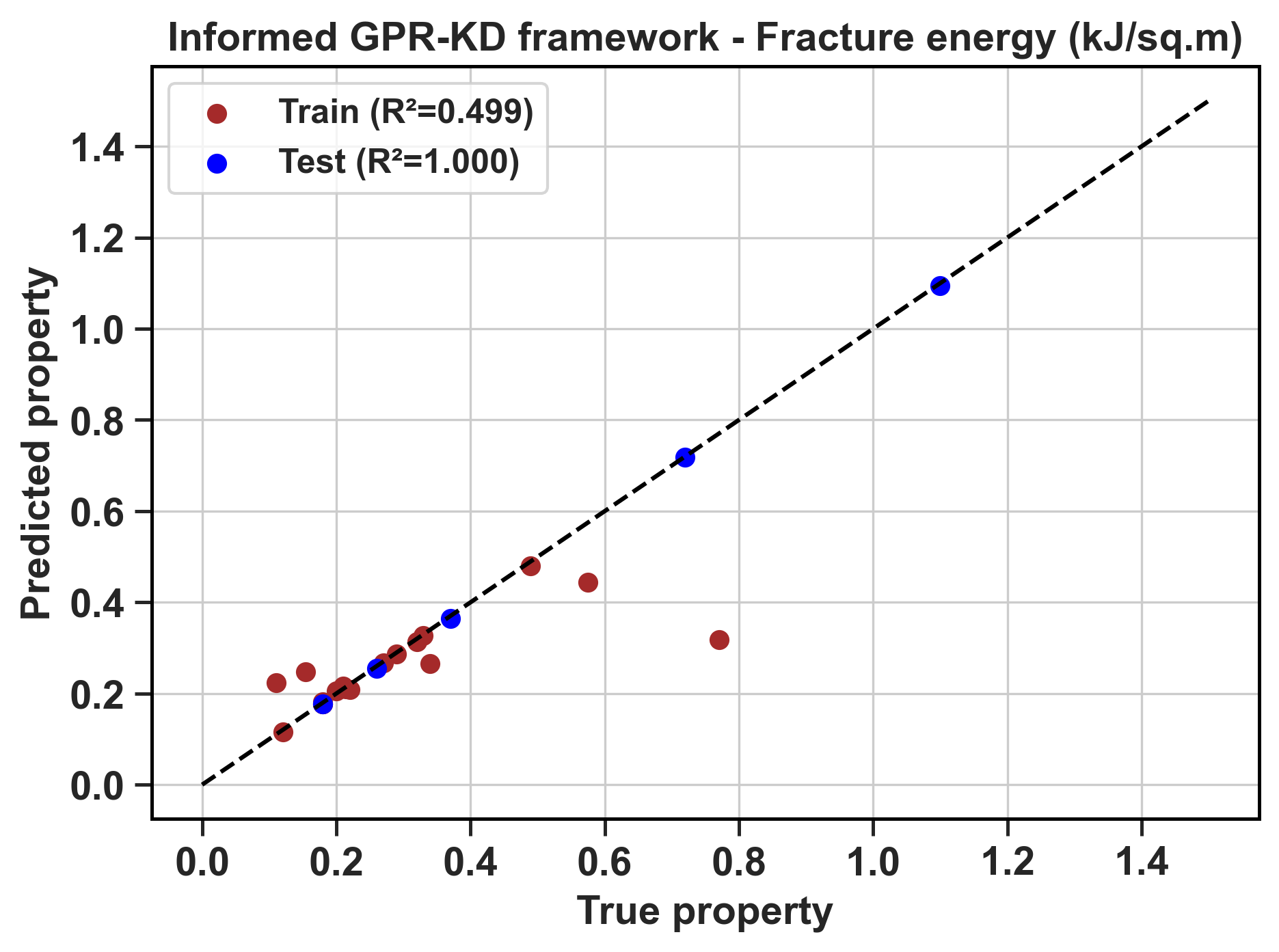}
        \caption{}\label{fig:6g}
    \end{subfigure}
    \begin{subfigure}{0.325\textwidth}
        \includegraphics [width=0.99\textwidth]{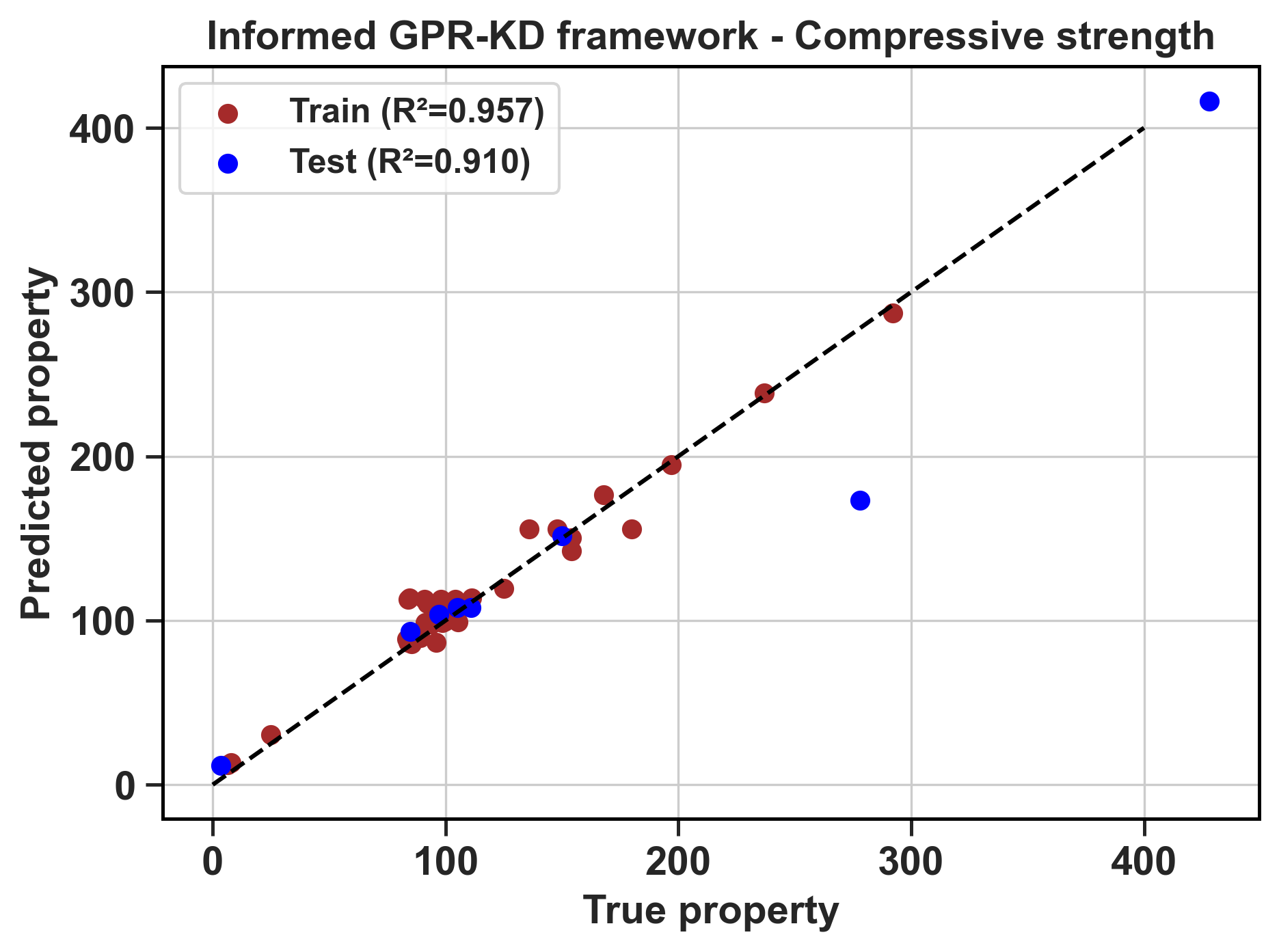}
        \caption{}\label{fig:6h}
    \end{subfigure}
    \caption{Comparison of true vs.\ predicted properties during simultaneous prediction of multiple properties using informed encoder-decoder model: (\subref{fig:6a}) Glass transition temperature $(^{\circ}$C); (\subref{fig:6b}) Density ($\mathrm{g/cm^3}$)); (\subref{fig:6c}) Adhesive strength ($\mathrm{MPa}$); (\subref{fig:6d}) Flexural strength ($\mathrm{MPa}$); (\subref{fig:6e}) Elastic modulus ($\mathrm{GPa}$); (\subref{fig:6f}) Tensile strength ($\mathrm{MPa}$); (\subref{fig:6g}) Fracture energy ($\mathrm{kJ/m^2}$); (\subref{fig:6h}) Compressive strength ($\mathrm{MPa}$)}
    \label{fig:6}
\end{figure}

The informed GPR-KD framework is then used for simultaneously predicting multiple physical and mechanical properties together. It is attempted to predict the properties like glass transition temperature, density, adhesive strength, flexural strength, elastic modulus, tensile strength, compressive strength and fracture energy all together using the informed GPR-KD framework. The prediction accuracy of all the properties (except compressive strength) improved while predicting them simultaneously as against individually predicting them (cf.\ Figure \ref{fig:5}). The improvement is primarily driven by the capability of the informed GPR–KD framework to capture common governing trends and correlated response patterns across different properties. By learning a unified feature space while predicting multiple properties simultaneously, the framework enables effective sharing of information among related properties. This simultaneous learning process constrains the solution space and acts as an implicit regularization mechanism, thereby reducing overfitting and improving the generalization performance of the model.
Finally, Figure \ref{fig:6} shows the comparison of predicted values with the true values for various physical and mechanical properties predicted simultaneously using the informed GPR-KD framework.  

\section{Conclusions} \label{sec:4}
In this study, attempt has been made to develop an informed Gaussian Process Regression based Knowledge Distillation (GPR-KD) framework for prediction of various physical (density and glass transition temperature) and mechanical properties (compressive strength, flexural strength, elastic modulus, tensile strength, adhesive strength and fracture energy) of thermoset epoxy polymers where the lack of availability of databanks/large curated datasets (as against general polymers) has limited the application of machine learning for these materials. Data from literature pertaining to experimental investigations on epoxy polymers covering wide range of resin/hardener classes and different physical and mechanical properties have been used for training the model. 
Following are some of the salient features of the proposed GPR–KD framework:
\begin{itemize}
    \item A hybrid teacher–student architecture in which Gaussian Process Regression (GPR) models serve as teacher models to guide a neural network student through knowledge distillation.
    \item GPR teachers capture nonlinear structure in limited data regimes and provide smooth, noise-robust predictions that act as informative soft targets for student training.
    \item Knowledge distillation enables the student network to inherit the generalization capability and regularization behavior of GPR while overcoming its scalability limitations.
    \item The framework explicitly incorporates characteristic descriptors of constituent resin and hardener molecules, allowing chemically meaningful learning of composition–property relationships.
    \item Separate embedding representations for resin and hardener constituents facilitate discrimination of similarly performing chemical systems and improve interpretability of learned trends.
    \item A single distilled student model is capable of predicting multiple physical and mechanical properties, eliminating the need for separate models for individual targets.
    \item Simultaneous multi-property learning promotes effective transfer of information across correlated properties, leading to enhanced predictive accuracy and stability.
    \item The GPR–KD framework demonstrates improved prediction accuracy compared to conventional standalone machine learning models, particularly in data-scarce regimes.
    \item By combining physics-informed regression with deep learning, the framework achieves a balance between interpretability, accuracy, and computational efficiency.
    \item The proposed approach is well suited for high-throughput screening and design of epoxy systems, enabling rapid exploration of composition–processing–property relationships.
\end{itemize}
This model thus paves the way for design of epoxy polymers with desired physical and mechanical properties thereby leading towards sustainable development.  

\section*{Acknowledgments}
The authors would like to thank the help received from Gregor Maier and Rick Oerder while developing the model. The first author (B.S.Sindu) acknowledges the support in the form of Post Doctoral Industrial Fellowship received from Indo-German Science and Technology Centre (IGSTC) to carry out research in Fraunhofer Institute for Algorithms and Scientific Computing (SCAI), Germany.

\printbibliography
\end{document}